# Concurrent Approach to Flynn's SPMD Classification

Through Java

Bala Dhandayuthapani Veerasamy
Department of Computing
Mekelle University
Mekelle, Ethiopia
.

*Abstract*—Parallel programming models exist as an abstraction of hardware and memory architectures. There are several parallel programming models in commonly use; they are shared memory model, thread model, message passing model, data parallel model, hybrid model, Flynn's models, embarrassingly parallel computations model, pipelined computations model. These models are not specific to a particular type of machine or memory architecture. This paper focuses the concurrent approach to Flynn's SPMD classification in single processing environment through java program.

## I. INTRODUCTION

Parallel programming and distributed programming [1] are two basic approaches for achieving concurrency with a piece of software. They are two different programming paradigms that sometimes intersect. In the past programming life, we were mostly using sequential programming. But, today's life style is going with more faster than the past decades. Also, solving problems on the computers are enormous. Parallel computer [1] can executes two or more job within a same period of time.

Two events are said to be concurrent if they occur within the same time interval. Two or more tasks executing over the same time interval are said to execute concurrently. Tasks that exist at the same time and perform in the same time period are concurrent. Concurrent tasks can execute in a single or multiprocessing environment [2]. In a single processing environment, concurrent tasks exist at the same time and execute within the same time period by context switching. In a multiprocessor environment, if enough processors are free, concurrent tasks may execute at the same instant over the same time period. The determining factor for what makes an acceptable time period for concurrency is relative to the application.

Concurrency techniques [3] [6] are used to allow a computer program to do more work over the same time period or time interval. Rather than designing the program to do one task at a time, the program is broken down in such a way that some of the tasks can be executed concurrently. In some situations, doing more work over the same time period is not the goal. Rather, simplifying the programming solution is the goal. Sometimes it makes more sense to think of the solution to the problem as a set of concurrently executed tasks. This technique is used in the parallel computer architectures.

Java is just a computer language [5] that has secure, portable, object-oriented, multithreaded [3] [4] [6], interpreted, byte-coded, garbage-collected, language with a strongly typed exception-handling mechanism for writing distributed programs [4]. Java is an object-oriented programming language, which added the new features such as overriding, interface and etc. Java supports multithreaded programming, which allows you to do many things simultaneously on the same time interval. Java enables the creation of cross-platform programs by compiling into an intermediate representation called java byte code. JVM (Java Virtual Machine) is an interpreter for java. Java is designed for the distributed environment on the Internet. Java has technology called RMI (Remote Method Invocation) that brings unparalleled level of abstraction to client / server programming. Byte code is a highly optimized set of instructions designed to be executed by the java run-time system, which is called Java Virtual Machine (JVM). Java handles de-allocation for you automatically, this technique called garbage collection. The Java Developers Kit (JDK) is a set of command-line tools that can be used to create Java programs. The current release of the JDK is version 1.6.

## II. FLYNN'S CLASSICAL TAXONOMY

Parallel computers can be divided into two main categories of control flow and data flow. Control-flow parallel computers are essentially based on the same principles as the sequential or von Neumann computer, except that multiple instructions can be executed at any given time. Data-flow parallel computers sometimes referred to as "non-von Neumann," is completely different in that they have no pointer to active instruction(s) or a locus of control. The control is totally distributed, with the availability of operands triggering the activation of instructions. In what follows, we will focus exclusively on control-flow parallel computers.

There are different ways to classify parallel computers. One of the more widely used classifications, in use since 1966, is called Flynn's Taxonomy [2]. The Flynn's taxonomy distinguishes multi-processor computer architectures according to how they can be classified along the two independent dimensions of Instruction and Data. Each of these dimensions can have only one of two possible states called Single or Multiple. There are four possible classifications according to Flynn's that is shown in Figure 1.

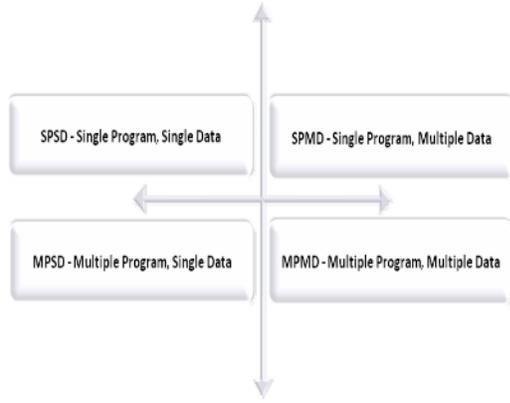

Figure 1. Flynn's Taxonomy

SPSD is simplest type of computer performs one instruction per cycle, with only one set of data or operand. SPSD is serial computer architecture. Such a system is called a scalar computer. SPSD will have one program and one set of data. Single instruction: only one instruction stream is being acted on by the CPU during any one clock cycle. Single data: only one data stream is being used as input during any one clock cycle. It will execute by only one processor as a sequential manner. Hence it is not a parallel programming model rather it is a sequential programming model. It can be executed by a single processor based system by sequential way.

MPSD [7] is a single data stream is fed into multiple processing units. Each processing unit operates on the data independently via independent instruction streams. Few actual examples of this class of parallel computer have ever existed. MPSD will have more than one program with same values will be used by all the programs. All the programs will execute in different processors with the same values. If all task or threads are executed in different processor means, it will take the same values.

SPMD is actually have a single program is executed by all tasks simultaneously. At any moment in time, tasks can be executing the same or different instructions within the same program. SPMD programs usually have the necessary logic programmed into them to allow different tasks to branch or conditionally execute only those parts of the program they are designed to execute. That is, tasks do not necessarily have to execute the entire program perhaps only a portion of it. Here all tasks may use different data.

MPMD [8] is actually a "high level" programming model. MPMD applications typically have multiple executable object files (programs). While the application is being run in parallel, each task can be executing the same or different program as other tasks. All tasks may use different data.

### III. FLYNN'S SPMD IMPLEMENTATION

SPMD is actually a "high level" programming model that can be built upon any combination of the parallel programming models. A single program can have multiple tasks, which can be executed simultaneously. At any moment in time, tasks can be executing the same or different instructions within the same program. SPMD programs usually have the necessary logic programmed into them to allow different tasks to branch or conditionally execute only those parts of the program they are designed to execute. That is, tasks do not necessarily have to execute the entire program - perhaps only a portion of it. Here all tasks may use different data.

Single processing environment can have concurrent tasks exist at the same time and execute within the same time period by context switching (time limits). This paper only focuses on Flynn's SPMD classification in single processing environment using concurrent approach.

Program 1. A Sample program for Flynn's SPMD

```
class SPMD implements Runnable{

Thread t;
String name;
int a,b,sum;

SPMD(String str, int val1,int val2){
    t=new Thread(this,str);
    name=str;
    a=val1;
    b=val2;
    t.start();  }

public void run(){
    try{

    sum=a+b; // single operation
    System.out.println("the sum is "+ sum + " produced
        by " + name +" thread");
    t.sleep(200);

    }catch(Exception e){ }
}

public static void main(String BDP[]){

SPMD b1=new SPMD("task1",1,1); // value 1
SPMD b2=new SPMD("task2",5,5); // value 2
SPMD b3=new SPMD("task3",10,10); // value 3
SPMD b4=new SPMD("task4",1,5); // value 4

}}
```

## IV. RESULTS AND DISCUSSION

Sequential programming also called serial programming. It is normally a computer programs, which involves consecutive process or sequential process. It can uses only single processing environment. There are drawbacks in sequential programming. It can be executed in a sequential manner. It will take more time for execution because instruction will execute by the processor one after another. It will have less clock speed. Biggest problems cannot be solved. The concurrent programming is also a computer program, which involves more than one process within the same time interval. Two or more tasks executing over the same time interval are said to execute concurrently. Parallel and distributed programmings are two approaches for achieving concurrency.

Java provides Thread class to create concurrent execution of tasks. Thread class has constructors and methods, which are helping to create concurrent execution of tasks. The Program 1 developed to execute in a single processing environment using Flynn's classification with SPMD. In this program, SPMD (String str, int val1, int val2) constructor has an argument with "String str" will receive name of the task, "int val1, val2" will receive values for addition, subtraction, multiplication and division. In the main function b1, b2, b3, b4 are objects; once it has created constructor will be automatically called. The t=new Thread(this, str) creates tasks such as "Task1", "Task2", "Task3" and "Task4". The t.start() method has been starting all the tasks. After starting the tasks, it will be automatically called run() method. The run() method will execute all the tasks concurrently by using t.sleep(200) method (context switching). The t.sleep(200) method pause any task for 200 milliseconds during execution and it will allow to execute waiting of any other task. Until completing all the tasks, run() method permitted to execute tasks concurrently using sleep() method and it enabled to exploit the processor idle time**.** Here all the tasks will use same operation with sum=a+b as the same program but it will use different data as "task1" uses 1,1; "task2" uses 5,5; "task3" uses 10,10; "task4" uses 1,5;  Finally, this program will produce the following result.

the sum is 2 produced by task1 thread

the sum is 10 produced by task2 thread

the sum is 20 produced by task3 thread

the sum is 6 produced by task4 thread

Hence, this research finding introduced to use for Flynn's SPMD classification to execute on the single processing environment using concurrent approach.

## V. CONCLUSION

Flynn's classical taxonomy has four classifications, except the first classification all other classification will utilize multiprocessing environment. While executing a single program over processor, it executes one task at a time. Concurrent approaches can execute multiple tasks using context switching in single processor environments. It enabled to execute Flynn's SPMD classification in single processing environment. This finding also initiated to have high performance and throughput on single processing environment. Hence, this paper recommending you to have concurrent execution of single program with multiple values on single processing environment.

AUTHORS PROFILE

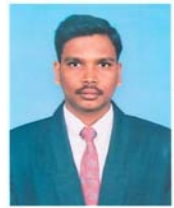

**Bala Dhandayuthapani Veerasamy** was born in Tamil Nadu, India in the year 1979. The author was awarded his first masters degree M.S in Information Technology from Bharathidasan University in 2002 and his second masters degree M.Tech in Information Technology from Allahabad Agricultural Institute of Deemed University in 2005. He has published more than fifteen peer reviewed technical papers on various international journals and conferences. He has managed as technical chairperson of an international conference. He has an active participation as a program committee member as well as an editorial review board member in international conferences. He is also a member of an editorial review board in international journals.

He has offered courses to Computer Science and Engineering, Information Systems and Technology, since 8 years in the academic field. His academic career started in reputed engineering colleges in India. At present, he is working as a Lecturer in the Department of Computing, College of Engineering, Mekelle University, Ethiopia.  His teaching interest focuses on Parallel and Distributed Computing, Object Oriented Programming, Web Technologies and Multimedia Systems. His research interest includes Parallel and Distributed Computing, Multimedia and Wireless Computing. He has prepared teaching material for various courses that he has handled. At present, his textbook on "An Introduction to Parallel and Distributed Computing through java" is under review and is expected to be published shortly. He has the life membership of ISTE (Indian Society of Technical Education).